\documentclass[12pt]{article}
\usepackage{xpatch}
\usepackage{times}
\usepackage{graphicx}
\usepackage{caption}
\usepackage{subcaption}
\usepackage{hyperref}
\usepackage{xcolor}
\usepackage{bibunits}
\usepackage{geometry}
\usepackage[normalem]{ulem}
\usepackage{bm}
\usepackage{amsmath,authblk}
\usepackage{setspace}
\def\arcsec{{$^{\prime\prime}$}}
\newcommand{\bcr}{\bf\color{red}}
\title{Spinning solar jets explained through the interplay between plasma sheets and vortex columns}
\author[1,2,3]{Sahel Dey}
\author[2]{Piyali Chatterjee}
\author[4,5,6]{Robertus Erdelyi}
\affil[1]{School of Information and Physical Sciences, University of Newcastle, University Drive, Callaghan, NSW 2308, Australia}
\affil[2]{Indian Institute of Astrophysics, Bangalore-560034, India}
\affil[3]{Joint Astronomy Programme and Department of Physics, Indian Institute of Science,
Bangalore-560012, India}
\affil[4]{Solar Physics and Space Plasma Research Centre (SP2RC), School of Mathematics and Statistics, University of Sheffield, Hicks Building, Hounsfield Road, Sheffield, S3 7RH, UK}
\affil[5]{Department of Astronomy, Eötvös Loránd University, 1/A Pázmány Péter sétány, H-1117 Budapest, Hungary}
\affil[6]{Gyula Bay Zoltán Solar Observatory (GSO), Hungarian Solar Physics Foundation (HSPF), Petőfi tér 3., Gyula, H-5700, Hungary}
\date{}
\begin{document} 
\maketitle
\iffalse
\author{Sahel Dey$^{1,2,3}$, Piyali Chatterjee$^{2}$, and Robertus Erdelyi$^{4,5,6}$}
\begin{affiliations}
\item School of Information and Physical Sciences, University of Newcastle, University Drive, Callaghan, NSW 2308, Australia
\item Indian Institute of Astrophysics, Bangalore-560034, India
\item Joint Astronomy Programme and Department of Physics, Indian Institute of Science,
Bangalore-560012, India
\item Solar Physics and Space Plasma Research Centre (SP2RC), School of Mathematics and Statistics, University of Sheffield, Hicks Building, Hounsfield Road, Sheffield, S3 7RH, UK
\item Department of Astronomy, Eötvös Loránd University, 1/A Pázmány Péter sétány, H-1117 Budapest, Hungary
\item Gyula Bay Zoltán Solar Observatory (GSO), Hungarian Solar Physics Foundation (HSPF), Petőfi tér 3., Gyula, H-5700, Hungary
\end{affiliations}
\fi
% Include the date command, but leave its argument blank.
\date{}
%%%%%%%%%%%%%%%%% END OF PREAMBLE %%%%%%%%%%%%%%%%


\baselineskip24pt

\maketitle 
\begin{abstract}
Bunches of swaying and spinning plasma jets in the solar atmosphere - the spicules- exhibit a variety of complex dynamics that are clearly observed in the images of the solar limb. 
Utilizing three-dimensional radiative magnetohydrodynamics (rMHD) simulation data, we uncover another facet of a forest of spicules that turns out to be a manifestation of the two-dimensional plasma drapery, instead of one-dimensional conical spikes. This fluted morphology is observed in other contexts like molecular clouds, auroras, and coronal loops. Further, using a sequence of high-cadence line-of-sight integrated images, generated from our simulation, we obtain  
multiple episodes of spinning amongst clusters of synthetic spicules, also reported in observations near the solar limb. This perception of rotation, according to our findings, is associated with hot swirling plasma columns, extending to coronal heights -- that we label as coronal swirling conduits (CoSCo). 
\end{abstract}

The dynamic complexity of solar spicules, discovered in 1877\cite{secchi1877}, is evident from several high-resolution observations of transverse oscillations\cite{pontieu12, kuridze16,bate22}, and bulk spinning motions\cite{suematsu08b} during their rapid evolutionary phase. 
%{\bcb Spicules are mainly observed in the chromospheric network and unipolar coronal hole regions. It is therefore argued that their forcing is hydrodynamic in nature, with the magnetic field simply providing a conduit for the jet\cite{Zirin88}}.
The observed lateral swaying is often interpreted as a manifestation of propagating fast magnetohydrodynamic (MHD) kink mode through spicules, while the spinning, though still debated, is believed to be either due to torsional Alfv\'en waves\cite{Sekse13, antolin18}, or because of mini-filament eruption\cite{sterling10b} or due to emerging new magnetic flux reconnecting over an active region\cite{Danilovic23}. %Images of the solar limb taken in the Ca II H line, from the Broadband Filter Imager (BFI) instrument on-board the Hinode spacecraft are dominated by swaying  spicules, believed so far to be cone-shaped jets. 
By analysing a sequence of data cubes at a high cadence of 2\,s from a series of high spatial resolution rMHD simulation of the solar atmosphere, we detect several spinning clusters of spicules that are often reported in solar atmospheric images taken in the Ca II H line from the Broadband Filter Imager (BFI) instrument on-board the Hinode spacecraft\cite{rompolt75,pishkalo94,suematsu08b,zaqarashvili09}. Apart from the rotation, sometimes a spicule is seen splitting into multiple strands. The reverse, where several strands in a bunch appear to join together into a single spicule during the course of spinning, is also detected in observations of the solar limb\cite{suematsu08b, sterling10a}. Here, we present a simultaneous explanation for both of the above observed phenomena, namely the spinning and the splitting of spicule strands with the help of an rMHD simulation including a 5\,Mm thick convection layer near the bottom of our box-in-a-star domain.
%With an imposed vertical magnetic field of 
%$B_{0}=7.5$\,G, reminiscent of the large scale polar field of the %Sun, in a gravitationally stratified medium. The convective turbulence acts on this imposed field to produce photospheric magnetic fields with strengths $\sim500$\,G by twisting and stretching motions of the turbulent plasma blobs. The coronal magnetic field strength, however, remains close to the imposed vertical field. No additional photospheric flux emergence is included in these simulations, such that the conditions resemble coronal holes. 

\subsection*{Spicules as warped plasma sheets.} %Forest of spicules are observed in several chromospheric and near transition region channels such as Ca II, Mg II, and Si IV owing to their multi-thermal structure (10,000 -- 80,000 K).
\begin{figure}
\centering
\includegraphics[width=0.69\textwidth]{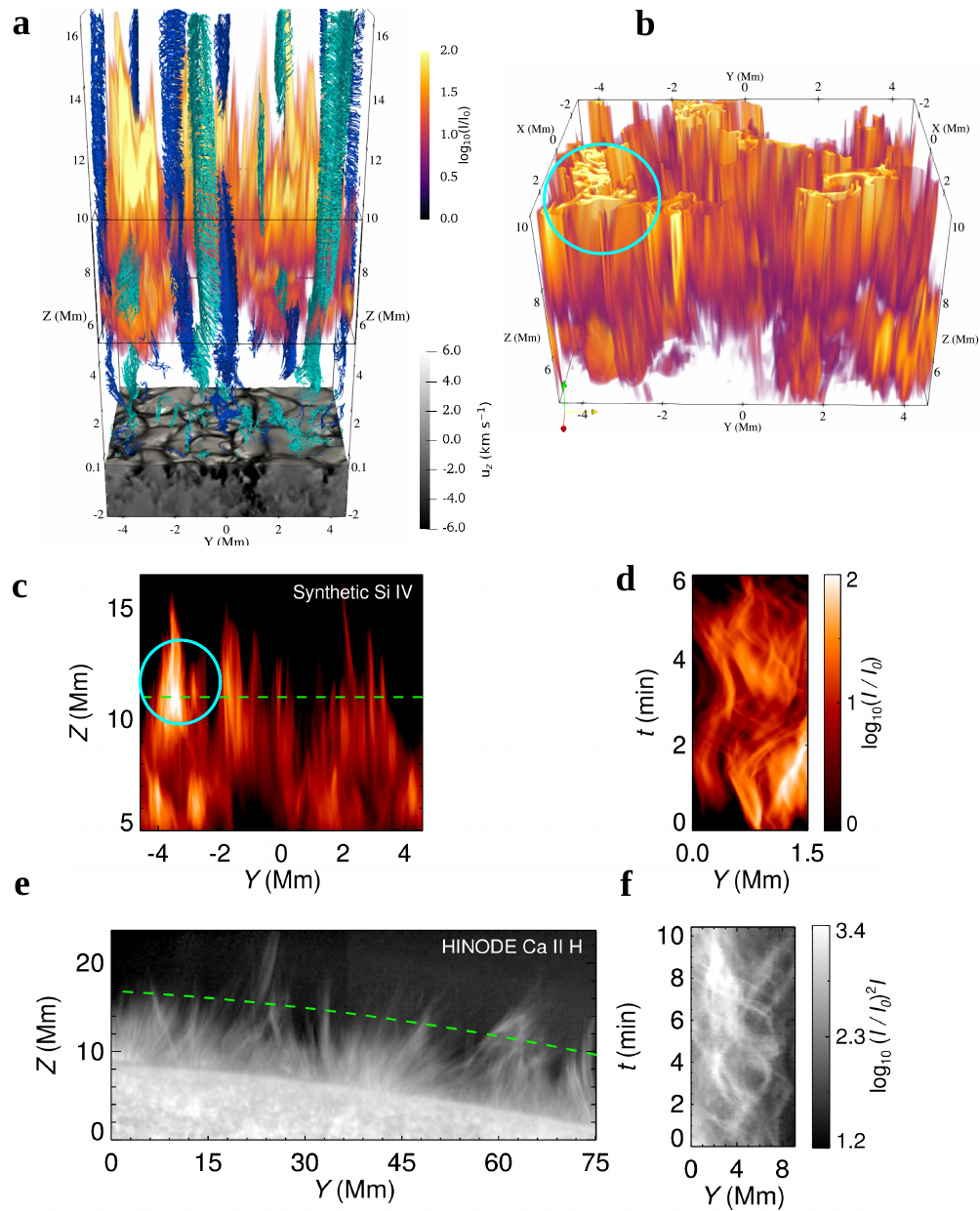}\caption{\label{fig:fig1}{Spinning cluster of solar spicules from simulation and observation. (a) Volume rendering of synthetic plasma emission at 80000\,K (similar to Si IV channel) with orange iso-contours and streamlines of CoSCos with both clockwise (cyan) and anti-clockwise (blue) rotational sense. Part of the turbulent convection zone (up to the visible photosphere) is shown as a grey scale block, colour coded by vertical plasma velocity. (b) Same as panel (a) but with a smaller subvolume (black box of panel a), showing the pleated curtain like morphology of spicules. The ratio of intensity of the synthetic emission between brightest to dimmest region of the visible sheet is a factor of 10. (c) Line-of-sight integration profile of same set of spicules of panel 
(b). The bright region encircled by the cyan circle has been obtained by an LOS integration over the densely pleated region of the same panel, also indicated by a cyan circle. (d) Time-distance diagram of synthetic spicules at $z=11$\,Mm, using a horizontal slit (green-dashed line of panel c). (e) SOT--Ca II H filtered image of solar spicules at 19:11:37 UT on 2007 November 7. (f) Time-distance diagram of spinning spicules at 8\,Mm height in the solar atmosphere, using a slit parallel to the limb (green-dashed line of panel e).}}
\end{figure}

In any snapshot taken of the three-dimensional simulation data, the spicules appear as pleated drapery of compressible plasma. This is
in contrast to a tube-like spicule geometry as evident in a volume rendering of synthetic emission from plasma at 80000\,K (Fig.~\ref{fig:fig1}a--b), comparable to the optically thin Si IV UV filter (see Methods) that is a solar transition region channel.  To make the pleating visually clear, a subdomain of Fig.~\ref{fig:fig1}a sliced at $z=10$\,Mm is shown in panel (b) and the Supplementary Video~\ref{mov:drapery_fig1} for further visual support. In the animation, we clearly see the plasma curtain rising and falling with a lifetime of 5--10\,min, commensurate with observations. Recent laboratory fluid experiments of Faraday excitation also point to such 
morphology\cite{Dey22}. The multi-threaded appearance of spicules, we find, is the result of a line-of-sight (LOS) integration of this dynamic drapery-like plasma sheet. The brightness of this drapery of plasma is highly non-uniform as seen in synthetic Si IV/80000\,K emission, particularly enhanced in the regions with several pleats. The spikes that appear brighter than ambient in the LOS integration therefore takes into account several folds of plasma (encircled by ellipses in Figs.~\ref{fig:fig1}b-c). The data from our three dimensional numerical simulations further self-consistently mimic most of the observed dynamic features of spicules. The spike-like LOS integrated jets in our simulations have the characteristics of observed spicules since these plasma features have a lifetime of 5-10\,mins, speeds of 20--70\,km\,s$^{-1}$, heights 6--16\,Mm, widths 200--1500\,km and their tips follow a parabolic path. This simulated forest of jets is excited in the presence of solar convection occurring in a layer below the photosphere and acting on a vertical magnetic field of magnitude $B_\mathrm{imp}=7.5$\,G imposed on the entire domain (see Methods). The convection excites slow MHD modes which form compression fronts (in blue) as they move into rarefied plasma as seen in  supplementary video~\ref{mov:drapery_fig1}. 
% that accelerate the spicule plasma at successive layers to move upwards in the first place.
These horizontally extended fronts, propagating upwards at the local sound speed are followed by the rising plasma drapery. A fluted sheet-like morphology for spicules has already been conjectured but not established\cite{Judge11,Judge12, Lipartito2014, Sterling20}, although the dense sheets reported here are not tangential discontinuities as assumed in the above works. In fact, as evident from the Supplementary Video~\ref{mov:spic_z9}, we find that the plasma sheet, as shaded by synthetic Si IV emission intensity, may have a variable thickness, varying between 100-600\,km. Similarly, a coronal veil hypothesis has also been put forward for overlying bright coronal loops above a simulated emerging active region\cite{Malanushenko22}. In the astrophysical context, the sheet-like morphology of 3D plasma structures is found in hydrogen molecular clouds in the galaxy that are regions of star formation, whereas in 2D images, they appear only filamentary\cite{Zucker21, Rezaei22, Dharmawardena23} Similarly, and as Earth's upper atmospheric analogue and counterpart, high-resolution images of auroras observed above the Earth’s atmosphere also show a fluted sheet-like morphology. %Synthesized H$\alpha$ movies of on-disk chromospheric fibrils from a very similar rMHD simulation but including rapid photospheric magnetic flux emergence also has appearance of a crumpled veil\cite{Danilovic22}.} 
Apart from spicules, other solar features like surges\cite{Nobrega18} and blinkers\cite{Nelson13} are also reported to display a filamentary appearance.

\begin{figure}
\includegraphics[width=\textwidth]{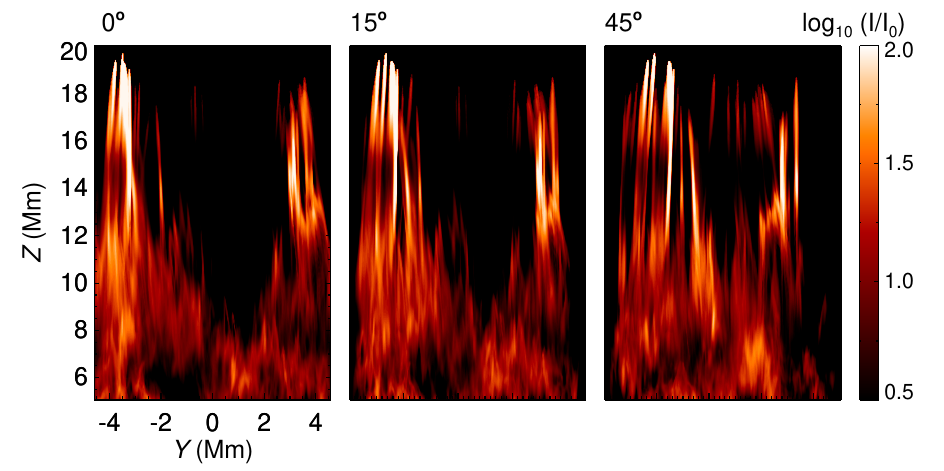}
\caption{\label{fig:extfig4}Simulated forest of spicules as seen from different LOS angles. LOS integration is performed along the (a) $x$-axis of the domain, (b) $15^{\circ}$ and (c) $45^{\circ}$, to the $x$-axis, respectively, to demonstrate how the multi-threaded structure and apparent synthetic intensity of spicule tips change with the viewing angle.}
\end{figure}

In order to quantify the transverse spicular motion, we present the LOS integrated synthetic intensity using a horizontal slit as shown in Fig.~\ref{fig:fig1}c. The appearance of inter-crossing features in the time-distance domain (panel d), spanning the duration of the Supplementary Video~\ref{mov:spic_rot}, is a clear signature of multiple threads crossing each other. The crossing can either be due to two neighboring threads each undergoing an independent transverse kink oscillation or due to spinning. We combine this with the animation of the volume rendering of the plasma drape (Supplementary Video~\ref{mov:drapery_fig1}) showing signs of 
stretching and curling when seen from an angle, to conclude that the spicules are indeed spinning and not just undergoing independent kink motions. Further, the LOS integration of volumetric plasma emission along different angles to the $x$-axis show that spinning multiple strands are actually different regions of the same dynamic plasma sheet. This can present an impression of splitting and merging to an observer along the $x$-axis at different epochs of the spinning motion (Fig.~\ref{fig:extfig4}). Yet another alternate explanation for the splitting and merging of spicules is based on complex Doppler variations in 2.5D simulations\cite{Pereira16}, even though it cannot rule out sheet-like structures.

\begin{figure}
\includegraphics[width=0.90\textwidth]{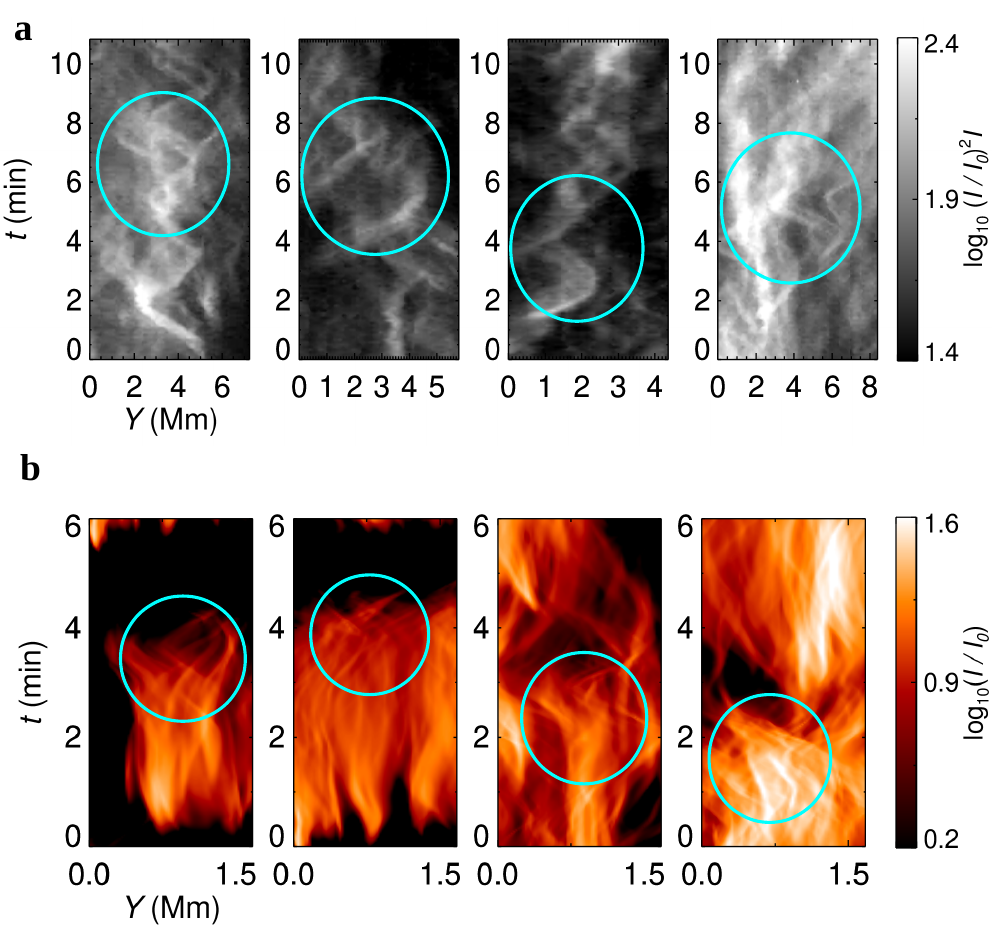}
\caption{\label{fig:extfig1}Identification of spinning spicules in observations and simulations using time-distance diagrams. (a) Scaled Ca II H/SOT intensity as seen through four different horizontal slits placed $z=8$\,Mm above and parallel to the observed solar limb similar to Fig.~\ref{fig:fig1}f to capture several examples of spinning clusters. Encircled regions (cyan) show crossing of spicule strands, a feature of spinning clusters of spicules. (b) Similar to panel (a), but for synthetic spicules seen through a slit at $z=11$\,Mm at four different times.}
\end{figure}

To compare and put in context with our rMHD simulation data, we analyse an observed dataset of 111\arcsec$\times$111\,\arcsec images of a region near the northern polar coronal hole taken on 07 Nov, 2007\cite{Judge10} by the Broadband Filter Imager (BFI) of Solar Optical Telescope (SOT) instrument\cite{suematsu08a} on-board Hinode\cite{kosugi07}. The images have been suitably filtered to increase the contrast of the spicule component (see Methods) as shown in Fig.~\ref{fig:fig1}e. A 56\,min long observation sequence is used when the instrument was pointing towards the limb. Five instances, where the bulk spinning motion is most discernible during the observation duration, are indicated in the Supplementary Video~\ref{mov:hinode_mov}, although more cases exist. 
The horizontal extents subtended on the plane of the sky by the spinning bunches vary between 2--4\,Mm. 
For each of these cases, we construct a time-distance map as seen through a slit placed locally parallel to the solar limb. One such case, shown in Fig.~\ref{fig:fig1}f and the other four examples in Fig.~\ref{fig:extfig1}a, can be compared with panel (b) for the bunch of simulated spicules. Independently, data from Si IV emission line images from IRIS\cite{IRIS14} taken on 19 Oct 2016 at 9.3\,s cadence also show clusters of spicules rotating in the second part of the Supplementary Video~\ref{mov:hinode_mov}.
%SD: Can we rewrite above part as -> Another independent observation of the Si IV emission line at 9.3\,s cadence from the IRIS mission dated 19 Oct 2016 also shows several spinning clusters of spicules.
However, because of the spatial resolution and signal-to-noise ratio being lower compared to the Hinode Ca II H data with 5 s cadence, the rotational motion is less discernible. 

%If the Reynolds number corresponding to transverse velocity is $>80$\cite{kundu08}, then it can also lead to Alfv\'enic vortex shedding\cite{Gruszecki10} for sub-Alfv\'enic speeds. 
%Compressibility is not an essential condition here, and for any non-zero viscosity, the vorticity at the boundary will eventually diffuse across the domain, outside the cylindrical tube. 

\begin{figure}
\includegraphics[width=\textwidth]{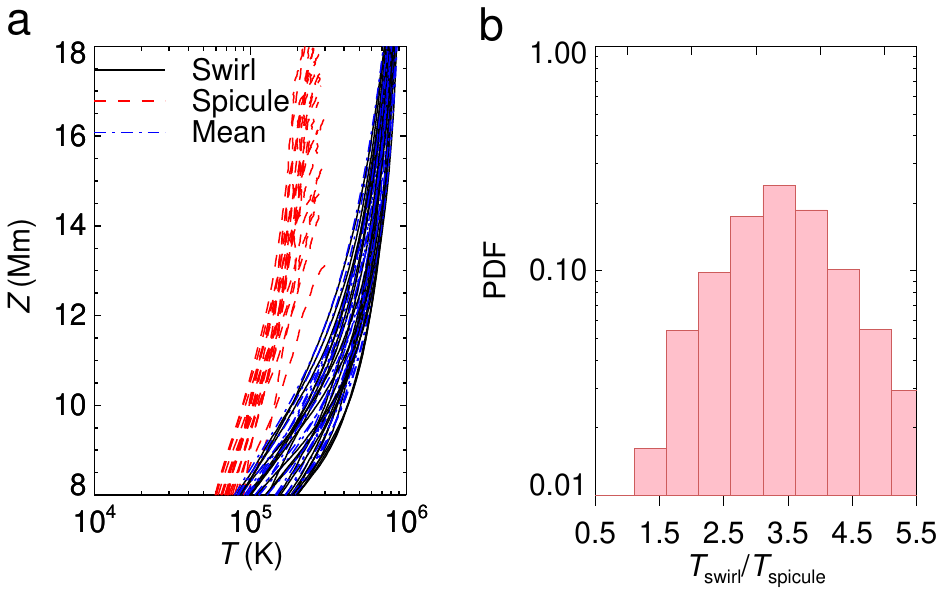}
\caption{\label{fig:temp}Temperature distribution of spicules and CoSCos, measured for a duration of 10 min of solar time. (a) Variation of horizontally averaged temperature profiles of CoSCos (black-solid), spicules (red-dashed) and the entire numerical box (blue-dashed dot) over different heights, computed at every 30\,s. (b) Number of grid points lying inside CoSCos boundaries, binned according to the temperature at those grid points, $T_\mathrm{swirl}$, scaled suitably to further support panel (a). The temperature at each such grid point has been scaled by the horizontally averaged temperature of spicules, $T_\mathrm{spicule}(z)$, at the height, $z$, corresponding to that grid point. To obtain the probability distribution function (PDF), the number of grid points inside CoSCos during a 10\,min duration or 20 time snapshots has also been scaled by the total number of grid points, $256 \times 384 \times 421 \times20$. Here, we have only taken CoSCos which do not overlap with regions with spicules, as defined using the contours of synthetic intensity.}
\end{figure}

In order to investigate the physical reason behind spinning spicule bunches reported in our simulation, we apply the Automatic Swirl Detection Algorithm (ASDA\cite{Liu19}, also see Methods) to simulation data. This technique is able to identify several highly dynamic, and helical velocity streamlines inside the 3D data cubes sampled at a cadence of 2\,s (see Supplementary Video~\ref{mov:3dspicule_swirl}). We introduce a new nomenclature -- coronal swirling conduits (CoSCo)-- for the stream tubes since they appear alongside spicules, are consistently tall cylindrical structures often much taller than spicules themselves and reach up to the corona ($>10$\,Mm), and have no significant connection to the photosphere, thereby differentiating them from previous work\cite{amari15,Finley22, Kuniyoshi23}, and a few other reported numerical simulations with the top boundary extending only till the upper chromosphere\cite{kitiashvili11, shelyag11,wedemeyer-bohm12,yadav21,Battaglia21}. Furthermore, note that with ASDA numerous CoSCos in the corona and far fewer in the chromosphere below are captured. Almost never did we find cases where the velocity streamlines defining CoSCos remain helical all the way down to the photosphere. This disconnect between the corona and the photosphere is highly suggestive of the hypothesis that these swirls may be generated in-situ in the solar atmosphere. The widths of CoSCos reported here are comparable (factor of 0.5) to spicule widths in the range 300--600\, km, rotational speed of 2--20\,km\,s$^{-1}${\bcr ,} and, have an average lifetime of 20--120\,s. Along its entire length, the swirl tubes are either clockwise or anti-clockwise. Approximately 75\% of all CoSCos detected in snapshots covering a duration of 10\,min are also noted to be hotter than the average temperature inside the spicules by a factor of 2.0--5.0 (Fig.~\ref{fig:temp}). 

\subsection*{Spicule--CoSCo interaction.} The CoSCos are seen forming both at the edges lining the spicules as well as outside. They in turn modify the surrounding flow propelling the spicules to rotate (Supplementary Video~\ref{mov:spic_z9}). The synthetic spicules are seen to rotate noticeably during their end-of-life. A spicule can rotate easily in its falling stage when most of the mass has drained towards the photosphere, compared to during the rising phase when the mass flux is upwards as evident from the 3D rendering in the Supplementary Video~\ref{mov:3dspicule_swirl}. Most of the CoSCos overlap only over a fraction of the entire lifetime of the nearby spicules. However, occasionally, we find swirls that rotate the adjacent spicules by moving inside them (e.g., the third part of Supplementary Video.~\ref{mov:spic_z9}). In such cases the CoSCos and spicules are co-spatial and their plasma properties cannot be distinguished.
%The analysis of wave modes \cite{khomenko18} shows presence of slow MHD mode compression fronts above the spicules (Supplementary Video~\ref{mov:slowm}). An analysis of time-periods of MHD slow mode in two randomly picked vertical cross-section of the 3D domain is given in the supplementary information file.
This kind of swirl generation, attached to spicules, exists in our simulation without any prior assumption about either the geometry of the spicule or the properties of its oscillation modes. A complementary mechanism involving the Lorentz force in which spicule matter can be lifted upwards by swirls has also been suggested\cite{Iijima17,scalisi21}.

In order to estimate the rotation velocity and lifetime of the spinning phenomena of CoSCos and spicules, the time-distance maps corresponding to spicule rotation from observations (simulations) are given in the top (bottom) row for several events in Fig.~\ref{fig:extfig1}. Notable differences are: i) that the number of thread crossings in the observed time-distance maps is more than in the synthetic case, ii) the time between two successive crossings is 1--4\,min in observations as compared to 0.5-- 2\,min in our simulations. This indicates that the spicule bunches take longer to spin but complete several spins during their lifetime on the Sun than in the simulations. The likely reasons could be %i) the spicule bunches are more coherent because of the imposed vertical magnetic field in the simulation domain, 
i) in the synthetic case, the CoSCos are seen forming near the end-of-life of spicules as the rotational energies are lower than observed, and ii) a larger computational domain size could support a bigger CoSCo that could not be captured by a horizontal extent of $6\times9$\,Mm$^{2}$. High spatial resolution observations in the blue wing of H$\alpha$ on the solar disk also show signatures of rapidly developing vortex-like features in spicules\cite{kuridze16}. 
That lateral motion of an initial dense cylindrical flux tube can induce vortices due to the Kelvin-Helmholtz instability (KHI) has been shown analytically\cite{barbulescu19} as well as using MHD simulations\cite{terradas08,antolin18}. The above scenario is similar to the case of a flow around a bluff body, where the velocity shear at a no-slip boundary causes boundary layer inversion and subsequently leads to the accumulation of small-scale vortices ($\ll$ size of the body) and a thickening of the boundary layer\cite{batchelor00}. The phenomena reported in this work is distinct from above, because we find folded sheets of stratified and highly compressible plasma instead of tube-like geometry of the spicules. In a follow-up study, we analyse and report the sources of vorticty for such in-situ generation of swirls and estimate the resulting upward Poynting flux. 

%Although many of the CoSCos form at the spicule periphery when the plasma is falling back to the chromosphere (see Supplementary Video~\ref{mov:spic_z9}), a few form while plasma is still shooting upwards. Such swirling conduits are directly associated with the spicules.
%We infer that the dominant vorticity sources are magnetic tension, $\left[\nabla \times (\bf{B}\cdot\nabla\bf{B}/\mu_0)\right]/\rho$ as well as baroclinicity due to combined effect of thermal and magnetic pressure, $\left[\nabla\rho\times(\nabla(p_g+B^2/2\mu_0)-\bf{B}\cdot\nabla\bf{B}/\mu_0)\right]/\rho^2$, acting mainly at periphery of the denser spicular plasma.

In summary, our simulations indicate a pleated drapery-like morphology for spicules in a coronal hole region on the Sun, without the inclusion of any photospheric flux emergence. The reason that spicules can still appear as conical jets in observations is because the sheets appear brighter only in regions where pleats are gathered, and, thus less bright parts of the plasma drapery remain below detection thresholds. The plasma sheets, whose thickness may vary from region to region, rise and fall with a lifetime commensurate with observations. Stereoscopic observations of the same region on the Sun performed simultaneously with a pair of telescopes for chromospheric or transition region lines may be required to establish whether spicules are morphologically two-dimensional fluted curtain embedded in three-dimensional plasma\cite{Judge11, Rast21} or simply one-dimensional straws. 
%Further, a domino-effect-like mechanism can explain a net positive mass flux found near the top of the domain and not just in the brightest regions of the plasma drapery (identified as straw-like spicules). The latter provides the mass reservoir for the fast solar wind along open magnetic field lines in the coronal hole regions\cite{Zirin88}.
Further, this morphological information would be crucial for estimating the mass and energy fluxes to the solar corona by considering the correct filling factor for solar spicules and not just the contribution from the brightest regions of the plasma drapery. 

\begin{figure}
    \centering   \includegraphics[width=0.6\textwidth]{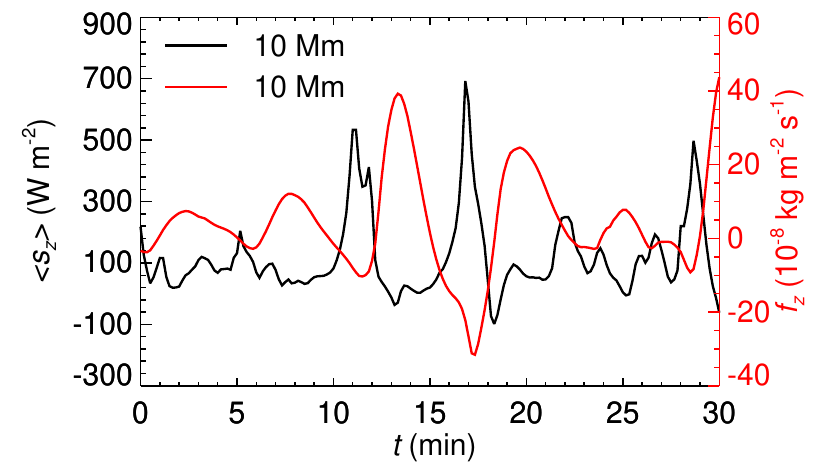}
    \caption{Horizontally averaged vertical Poynting flux, $\langle S_z\rangle=\langle v_zB_h^{2}-B_z(\vec{v}_h\cdot\vec{B}_h)\rangle$ (black) and mass flux, $f_z=\langle\rho v_z\rangle$ (red), over $6$\,Mm$\times 9$\,Mm, as a function of time calculated at $z=10$\,Mm. $\vec{v}_h$ ($v_z$) and $\vec{B}_h$ ($B_z$) are horizontal vector (vertical component) of velocity and magnetic field, respectively. }
    \label{fig:poynting_massflux}
\end{figure}

In contrast to spicules that fall back to the solar chromosphere, the CoSCos and the associated shearing flow form by drawing energy from the thermal and magnetic energy reservoir of the chromospheric plasma. A manifestation of this is evident by plotting the horizontally integrated Poynting flux, $\langle S_z\rangle(t)$, at coronal heights as a function of time. The $\langle S_z \rangle$ at $z=10$\,Mm is maximum (minimum) when the spicules are falling (rising). In other words, $\langle S_z \rangle$ and the horizontally integrated mass flux are anti-phase, as shown in Fig.~\ref{fig:poynting_massflux}, akin to the conversion between kinetic and potential energies in a simple harmonic oscillator. This is also supported by the Supplementary Video~\ref{mov:3dspicule_swirl} showing the presence (absence) of swirls (spicules) and vice versa at any given time. 
Therefore, CoSCos have the potential to carry the rotational energy further upward than the typical length of spicules seen in the Si IV emission lines even when spicules are falling sun-ward. 
Detection of the rotating bunches of spicules both in observations and in simulations presented here affirms that rotating spicules on the Sun mark regions where CoSCos exist. Therefore a complete picture of momentum and energy transport to the fast solar wind must take into account this coupling between spicules and CoSCos. 
%%High grid resolution simulations like reported here but with a significantly larger horizontal domain will be reported in near future to understand the statistical similarities and differences between morphology and energy transport by CosCos of both kinds.
Looking forward, observations with high resolution and sensitivity taken simultaneously in different layers of a locally unipolar magnetic region on the Sun\cite{Berger04} will help in investigating the energy exchange between spicules and swirls, and eventually its outward transport. %with e.g., Ca II K, Fe XI and H$\alpha$ channels, TiO and Fe XI are both photospheric lines although Fe XI can have some coronal part. 

%\begin{methods}
\subsection*{The radiative MHD set-up:}
We solve the equations of magneto-hydrodynamics (MHD) coupled with radiative transfer by extending a publicly available 2-dimensional setup spanning part of the solar convection zone to the lower solar corona\cite{chatterjee20} to 3-dimensions. The radiative transfer equation is solved using the Rosseland mean opacities for the optically thick solar atmosphere below $z=1.2$\, Mm, using the technique of long characteristics\cite{heinemann06}. For the optically thin plasma above, we use a cooling function, $Q_\mathrm{thin}=y_\mathrm{H}n_\mathrm{H}^2\Lambda(T)$, where $y_\mathrm{H}$ is the ionization fraction of Hydrogen, and $n_\mathrm{H}$ is the number density of the element Hydrogen. $\Lambda(T)$ is a function of temperature in units of J m$^3$s$^{-1}$ and available in tabulated form, computed from atomic data\cite{Cook89}. The radiation that escapes outward contribute to the cooling of the solar plasma. At the same time, an equal part of the optically thin radiative flux is emitted radially inward and is absorbed in the chromosphere, where it contributes to radiative heating instead of cooling\cite{Carlsson12}. To calculate this heating at every time-step in a 3-dimensional simulation is non-trivial, therefore we approximate  $Q_\mathrm{thin}\rightarrow 0$ smoothly for $T<30,000$\,K. 

We use the fully compressible higher-order finite difference MHD tool, the {\sc Pencil Code}\cite{brandenburg21} (hosted at \verb|https://github.com/pencil-code/|). This code uses sixth-order finite difference scheme for spatial derivatives and a three-step Runge-Kutta-Williamson time stepping scheme out of several other available modules. The {\sc Pencil code} includes several non-ideal MHD parameters, e.g., explicit viscosity and Ohmic 
resistivity, anisotropic Spitzer conductivity, shock viscosity and hyper-dissipation\cite{chatterjee20}. The domain extent is 6$\times$9\,Mm$^2$ in the horizontal direction. 
The bottom of the domain lies at $z=-5$\,Mm (in the convection region) and the top extends to 32 Mm height (the lower corona), above the photosphere located at $z=0$. The bottom boundary is open to conductive and radiative fluxes. To minimize the reflection of outgoing waves at the top boundary, we place a 6\,Mm thick sponge layer at $z=26$\,Mm to absorb any outgoing fluxes by adding damping terms to 
velocity equation. Further, in absence of any self-sustained heating process, the temperature in the sponge layer is maintained at $10^6$\,K~\cite{Rempel21,Kuniyoshi23}. Such a technique was used in 3D simulations earlier\cite{Iijima17, Kuniyoshi23}. We have a shallow convection zone and a primarily locally unipolar region. The constant temperature in the sponge layer has been included as a precautionary measure so as to not allow the atmosphere to collapse during the time the convection is building up. In any case, the sponge layer cannot heat the atmosphere to reach temperatures beyond million Kelvin. Our model also does not include any flux emergence and in the absence of it, the coronal temperature can cool to 
$< 200$kK\cite{Finley22}.
We impose a vertical magnetic field in the simulation box and gradually increase its strength from 0.25\,G to the final value of $B_{0}=7.5$\,G in 25\,minutes of solar time. This strength is reminiscent of the large scale polar field of the Sun. After evolving for $\sim$50 minutes, we observe several turbulent granulation cells on the photosphere.  
In the presence of turbulent convective flows, the magnetic field strength reaches $\sim300$--500\,G in some regions at the photosphere. These localized strong magnetic regions are known as Magnetic Bright Points (MBPs) and are found inside downflow lanes. No additional magnetic flux emergence process is employed at the bottom boundary. In our simulation, the plasma-$\beta$ remains $\sim0.05$ for $z>6$\,Mm. %For the run~4, we use a harmonic forcing at the bottom boundary instead of convection. A very shallow sub-surface layer 300\,km thick is used here. The applied membrane forcing is of the form $V_0 \sin(2\pi f_0 t)\cos(\pi x/L) \cos(\pi y/L)$, with $f_0=3.3$\,mHz and $V_0=1.32$\,km\,s$^{-1}$, and only applied within a layer, $-100$\,km $<z<0$\,km.
Here, we report results from simulations with a uniform grid resolution of 24 km with imposed vertical magnetic field of 7.5\,G. We also perform simulations with a coarser grid resolution of 48\,km and an imposed field of 3\,G and 25\,G (Run~1 and~2). Multitude of spicules are excited when the convective buffeting with a period $\sim$5\,min give impulsive kicks to the lighter chromospheric plasma\cite{DePontieu04}. The slow MHD waves thus excited, steepen into shocks during their upward propagation due to the strong density stratification above the solar photosphere. The density compression ratio is in the range 1.86--2.19 for $3.5<z<4.5$\,Mm for few representative shock fronts. Using the Rankine-Hugoniot jump condition, we estimate the corresponding upstream Mach number, $v/c_s \sim$ 1.61--1.90 with respect to the shock front reference frame. The heating inside the compression regions identified as shock fronts comes mainly from plasma compression, -$(\gamma-1)T \nabla\cdot\mathbf{v}$, with $\sim$ 1\% contribution coming from the dissipation of shock due to the explicit shock viscosity $\propto -\nabla\cdot \mathbf{v}$ in the code. Therefore, the shocks are approximately adiabatic (see the Supplementary information file). In the setup, in addition to explicit diffusion, we use a sixth-order hyper-dissipation scheme that conserves both the linear and angular momentum\cite{lyra17} for compressible plasma. We have explored a few numerical setups by selecting different spatial resolutions, domain size, strength of imposed magnetic field. In all the cases, key features e.g., forest of spicules, ubiquitous swirls, spinning clusters of spicules, remain robust. For the simulation runs presented here, we relax the solution for first 50 min starting from initial stratification at $t=0$. No additional wave damping was added to remove any global oscillations during the relaxation time. Then, we continue for a further 30 min and analyze this data. At this time the convection spans the entire depth of the convection zone and the root-mean-square value of velocity in the convection zone has saturated with time. The initial shocks due to lack of numerical equilibrium are not found after the relaxation period.
\newline
%The run directory for the 3D convective simulation is available as part of the sample section of the {\sc Pencil code} distribution at \verb|samples/solar_spicule3D|\footnote{The link will be available after the refereeing process.}, and includes the initial (running) parameters in the files \texttt{start.in (run.in)}, 
%the initial stratification file of density and temperature , and a reference file for comparison. Additionally, the compilation files \verb|src/Makefile.local| provide the list of incorporated physics modules in the setup, and \verb|src/cparam.local| sets the grid size as well as the domain decomposition over multiple CPUs, respecting the cache efficiency of the code.
\subsection*{Synthetic plasma emission:}
The Hinode images are taken in the chromospheric line (Ca II H at $T_\lambda \approx 15000$\,K).  This line is difficult to forward model due to partial re-distribution effects and strong departures from local thermodynamic equilibrium. Therefore, we choose instead to compute the synthetic intensity of the 80000\,K plasma emission which is optically thin and therefore easier to synthesize directly from MHD variables in our simulation. 
The expression for the synthetic intensity may be given by,
\begin{equation}
\label{eq:syn}
I_\lambda=\int Q_\lambda(\rho, T) \rho^2 ds,
\end{equation}
that hold for temperatures between $10^4$--$10^8$\,K\cite{Landi08}. The contribution function,  $$Q_\lambda(\rho, T)=\exp\left[-\left(\log(T/T_\lambda)/w\right)^2\right],$$ with $T_\lambda=80000$\,K (compare with optically thin line Si IV) and, $w=\log2.0$. The integration is performed along the line-of-sight with the infinitesimal path length denoted by $ds$. We have compared both the volume rendering and LOS integration of Si IV emission from Eq.~\ref{eq:syn} with FoMo \cite{Doorsselaere16} that uses the CHIANTI package for contribution functions of optically thin lines (see also Supplementary information file).  Spicules consist of multi-thermal plasma emitting at several wavelengths. Therefore, if the spicule structure undergoes certain dynamics, e.g., swaying or rotation, we would expect the corresponding emission to also reflect the same dynamics. The agreement between the images using an analytical contribution function and from FoMo shows that even if the calculation of the emission is not accurate, the time variation of the original feature will still be reflected in the forward modelled emission.

\subsection*{Vortex detection and visualization with ASDA and LIC}
The detection of vortices over multi-dimensional data is a active field of research and its usage can be found across different disciplines. There are several methods namely Automatic Swirl Detection Algorithm\cite{graftieaux01} (ASDA), Swirling strength criterion \cite{zhou99}, Lagrangian Averaged Vorticity Deviation (LAVD)\cite{haller16} to detect vortices over different time frames. Here, we use the ASDA method to track swirls distributed between the solar photosphere and lower corona. In this method, there exists a free parameter known as window size, which acts as local filter to smooth out the small scale fluctuations in the velocity field. We have selected a window of 480 km $\times$ 480 km, which is typical diameter of existing vortices. Over the window in each horizontal plane, two crucial quantities $\gamma_1$ and $\gamma_2$ are computed to locate center and circumference of the vortex tube. Incompressible fluid models provide a range of threshold values of axisymmetric vortices to demarcate core rotating region. In our compressible turbulent convection model, we have noticed that $|\gamma_1| \geq 0.55$ regions distinguish clockwise and anti-clockwise rotating non-axisymmetric vortices even in presence of strong shear flows. We have also confirmed these regions independently by comparison with a 2D flow visualization method known as Line Integral Convolution (LIC) applicable to streamlines in the horizontal plane. We apply this method at every horizontal layer in the height range 0--20\,Mm and track the vortex tubes over the time duration of 30\,minutes.

\subsection*{Filtering of Hinode SOT images:}
To enhance the intensity of spicules over the background, we employ radial filtering technique on the Level-1 Hinode Solar Optical Telescope (SOT) data sets, obtained after \verb|fg_prep.pro| (available under SolarSoftWare suite written in IDL) operation on the raw images as part of the calibration. We first compute the non-dimensional weight factor as the ratio of intensity at each pixel to the average intensity over all pixels lying on the arc parallel to the solar limb. After applying the same operation to all radii, we multiply intensity with the square of the weight factor at every pixel. This makes relatively brighter (dimmer) pixels to more brighter (dimmer) than before and enhances the visibility of the spicule forest compared to the surroundings. A similar method\cite{DePontieu07} was used earlier as well for off-limb spicules in Hinode data.
\subsection*{Processing IRIS Slit-Jaw images:}
We have selected an 82 Mm $\times$ 36 Mm region inside a coronal hole near the solar north pole with IRIS Slit-Jaw SI IV filter images, starting from 19 Oct 2016 at 19:04:09 UT. The analyzed observation sequence is 39\,min long. The level-2 calibrated images are first cleaned up by removing cosmic rays and dust spots by using \verb|array_despike.pro| and  \verb|iris_dustbuster.pro| (available under SolarSoftWare platform), respectively. The level-2 corrections on IRIS data can move values outside the range of 14-bit pixel values. Hence, it should be scaled appropriately using an algorithm that also applies a normalization based on exposure time to the image sequence. This kind of scaling allows maximizing the information content of the processed animation. Therefore, we scale the intensity from 14 bits to 8 bits pixel values with \verb|iris_intscale.pro| as suggested in the manual to enable us to clearly show the fine-scale structure of spicules. 
%\end{methods}

\section*{Data Availability}
The simulation data used to produce Fig.~\ref{fig:fig1} will be made available upon request. The Hinode SOT datasets used in this paper can be downloaded from the \href{https://darts.isas.jaxa.jp/solar/hinode/query.php?coA=basic&fgA=advanced&spA=basic&xrA=basic&eiA=basic&coB=&coC=&coD=0&coE=&coF=0&coG=1&coH=0&coI=0&coJ=0&coK=0&coL=0&coM=0&coN=0&coO=0&coP=0&coQ=0&spB%5B%5D=0&spC=&spD=1&spE=&spF=&spG=&spH=&spI=&spJ=&spK=&spL=&spM=&spN=&spO=&spP=&spQ=&spR=&spS=&spT=&spU=&spV=&xrB%5B%5D=0&xrC=&xrD=&xrI=&xrE=&xrF=&xrG=&xrH=&xrJ=&xrK=&xrL=&xrM=&xrN=&xrO=&xrP=&xrQ=&xrR=&eiB%5B%5D=0&eiC=&eiD=&eiM=&eiE=&eiF=&eiG=&eiH=&eiJ=&eiK=&eiL=&eiN=&eiO=&eiP=&eiQ=&eiR=&eiS=&A19=Search&inst%5B%5D=fg&plot=no&tmSY=2007&tmSM=11&tmSD=07&tmSh=18&tmSm=30&tmEY=2007&tmEM=11&tmED=07&tmEh=19&tmEm=30&posX=97&posY=939&posrX=111&posrY=111&fgB%5B%5D=0&fgD=&fgE=0&fgG=&fgH=&fgI=&fgJ=&fgL=&fgM=1&fgN=0&fgO=0&fgP=0&fgQ=0&fgR=&fgS=&fgT=}{DARTS} archive. The analyzed IRIS SI IV Slit-Jaw images are available at the \href{https://www.lmsal.com/hek/hcr?cmd=view-event&event-id=ivo://sot.lmsal.com/VOEvent%23VOEvent_IRIS_20161019_190409_3620109603_2016-10-19T19:04:092016-10-19T19:04:09.xml}{Heliophysics Events Knowledgebase}
%\verb|https://www.lmsal.com/hek/hcr|

\section*{Code Availability}
The Pencil code is open source and hosted at \verb|https://github.com/pencil-code/|.

\clearpage
\bibliography{reference}
\bibliographystyle{naturemag}
\clearpage
\clearpage
\begin{singlespacing}

\end{singlespacing}
\clearpage
%%%%%%%%%%%%%%%%%%%%%%%%%%%%%%%%%%%%%%%%%%%%%%%%%%%%
%%%%%%%%%%%%%%%%%%%%%%%%%%%%%%%%%%%%%%%%%%%%%%%

\setcounter{figure}{0}
\begin{figure}
\renewcommand{\figurename}{Video}%
    \caption{%\href{https://drive.google.com/file/d/1XMq5upai7wUkeS_CZu-3YcdFyUgT30BV/view?usp=sharing}
    {\label{mov:drapery_fig1} Animation for the volume rendering of the spicule drapery presented in Fig.~\ref{fig:fig1}b. Slow mode shock fronts (indicated by blue shaded regions) accelerate the spicule plasma upward. The slow mode amplitude is calculated using $\mathbf{\hat{n}.\nabla(\hat{n}.v)}$, where $\mathbf{\hat{n}}$ is a unit vector parallel to the magnetic field.}}
\end{figure}

\begin{figure}
\renewcommand{\figurename}{Video}%
\caption{%\href{https://drive.google.com/file/d/1EJRAF57LW0eXdjBZilwBovRcYrtR733l/view?usp=sharing}
{\label{mov:spic_rot} Synthetic spicules and their spinning motion modeled using LOS integrated plasma emission at 80,000 K for different Field of Views (FOVs) from the numerical experiment.}}
\end{figure}
\begin{figure}
\renewcommand{\figurename}{Video}%
\caption{%\href{https://drive.google.com/file/d/1ZNr0EDLMeKHAMrgFgQu25VjRLzg0Rmbu/view?usp=sharing}
{\label{mov:hinode_mov}({\em First part}) Rotating bunches of spicule strands observed at the Solar limb by high cadence Hinode SOT BFI images (plasma emission at 15,000 K) for a duration of 41 min. ({\em Second part}) Similar but for a 39\,min long IRIS Si IV observation, which is sensitive to plasma temperature of 80,000 K}}
\end{figure}
\begin{figure}
\renewcommand{\figurename}{Video}%
\caption{\label{mov:3dspicule_swirl} Interaction of synthetic spicules (shaded yellow) with CoSCos (counterclockwise rotating ones represented by red streamlines) shown in 3-dimensions. The animation covers 10\,min of solar time. Here, one can note that CoSCos typically form either around falling spicules or in regions where there is reduced spicular activity.}
\end{figure}

\begin{figure}
\renewcommand{\figurename}{Video}%
\caption{%\href{https://drive.google.com/file/d/18jT-Avg0FIH1xb-gdglbgPjjk8uRF1Jo/view?usp=share_link}
{\label{mov:spic_z9} Interaction of CoSCos with synthetic spicules result in their spinning motion. A cross section of synthetic intensity of spicules at $z=9.0$ Mm is shown in the horizontal plane and plasma flow is visualized through Line Integral Convolution (LIC) technique.}}
\end{figure}
\end{document}